\documentclass[aip, apl, reprint,twocolumn]{revtex4}
\usepackage{amsmath,amssymb}

\usepackage{graphicx}
\usepackage{dcolumn}
\usepackage{bm}
\usepackage{hyperref} 

\begin{document}

\title{In-depth analysis of CIGS film for solar cells, structural and optical characterization}

\author{A.~Slobodskyy$^{1,2}$}
\email[]{anatoliy.slobodskyy@kit.edu} 
\author{T.~Slobodskyy$^{3}$}
\author{T.~Ulyanenkova$^{3}$}
\author{S.~Doyle$^{3}$}
\author{M. Powalla$^{2}$}
\author{T.~Baumbach$^{3}$}
\author{U.~Lemmer$^{1}$}
\affiliation{$^1$Light Technology Institute, Karlsruhe Institute of Technology (KIT), Kaiserstr. 12, 76131 Karlsruhe, Germany\\
$^2$Zentrum f\"{u}r Sonnenenergie- und Wasserstoff-Forschung Baden-W\"{u}rttemberg, Industriestr. 6, 70565 Stuttgart, Germany\\
$^3$Institute for Synchrotron Radiation, Karlsruhe Institute of Technology (KIT),76344 Eggenstein-Leopoldshafen, Germany}

\date{\today}

\begin{abstract}
Space-resolved X-ray diffraction measurements performed on gradient-etched CuIn$_{1-x}$Ga$_x$Se$_2$ (CIGS) solar cells provide information about stress and texture depth profiles in the absorber layer. An important parameter for CIGS layer growth dynamics, the absorber thickness-dependent stress in the molybdenum back contact is analyzed. Texturing of grains and quality of the polycrystalline absorber layer are correlated with the intentional composition gradients (band gap grading). Band gap gradient is determined by space-resolved photoluminescence measurements and correlated with composition and strain profiles.
\end{abstract}


\maketitle

High-efficiency thin-film solar cells based on CuIn$_{1-x}$Ga$_x$Se$_2$ (CIGS) films attracted major interest as a source of renewable energy. Flexibility of the CIGS quaternary alloy boosts scientific interests to improve the solar cell performance. The band gap of CIGS can be changed with the composition defined by the Ga/(Ga+In) (GGI) ratio. A band gap gradient within the solar cell absorber layer is used to improve solar cell efficiencies~\cite{Contreras1999}. The lattice constant, strain in the layer, average grain size and device properties are influenced by the composition of the chalcopyrite~\cite{Abou-Ras2008}. The substrate induced strain and band gap grading are shown to be important factors for  proper understanding of the absorber growth dynamics and solar cell performance overall. In this contribution we present a detailed experimental analysis of the strain profile within a CIGS solar cell absorber layer in a high-efficiency solar cell.

\begin{figure}
\includegraphics{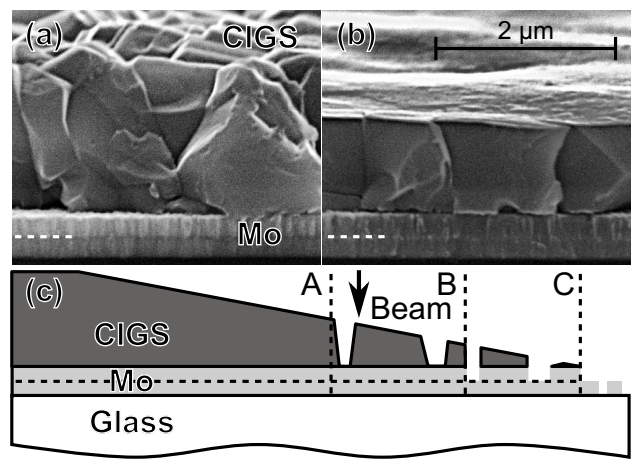}
\caption{\label{sample} Cross section of samples used in the experiments. (a) SEM image of polycrystalline CIGS absorber on top of the Mo layer. (b) SEM  image of etched CIGS layer cross section showing a planarization of the layer top surface. (c) Drawing of the sample with etched gradient of the absorber layer.}
\end{figure}

The samples used in the experiments were prepared from solar cells fabricated using a modified co-evaporation process. The solar cell absorber layer was evaporated on a soda lime glass substrate covered with 500~nm of a Mo-sputtered in two steps back contact. The absorber thickness was 2.2~$\mu m$. Nearly 500~nm of ZnO:Al top contact and 80~nm of buffer CdS were removed by a selective etching process with hydrochloric acid. The cross section of the samples is shown in Fig.~\ref{sample}~(a).

The CIGS absorber layer was etched with aqua regia formed by mixing hydrochloric and nitric acid. An example of an etched absorber layer is shown in  Fig.~\ref{sample}~(b). The dissolving activity of aqua regia and its corresponding etching rate vary with its molar mixing ratio. A HCl to HNO$_3$ ratio of 1:3 was found to result in etching rates of approximately 10~nm/sec 3 minutes after preparation of the acid. The gradient in the absorber thickness along the sample plane is achieved by continuous withdrawal of the sample from aqua regia during etching. A schematic drawing of the sample with the absorber layer gradient is shown in Fig.~\ref{sample}~(c). We divide the etched sample in few regions depending on properties of etched layers. The regions are separated at positions~A, B and C shown by vertical dashed lines in Fig.~\ref{sample}~(c).

Synchrotron measurements were performed using a monochromatic beam of 10.8~keV at the P-Diff beamline at ANKA synchrotron facility. The samples were mounted on a four-circle diffractometer. Sample translation in the direction perpendicular to the beam propagation was performed using a motorized translation stage. 2D synchrotron X-ray powder diffraction images were collected at each sample position using a two-dimensional (2D) CCD detector with the pixel size of 60x60~$\mu m^2$ located 207~mm away from the center of the diffractometer. Using the collected data, we were able to map a region of $2\theta$ angles from 17 to 40~degrees and $\phi$ angles from 18 to 90~degrees.

\begin{figure}
\includegraphics{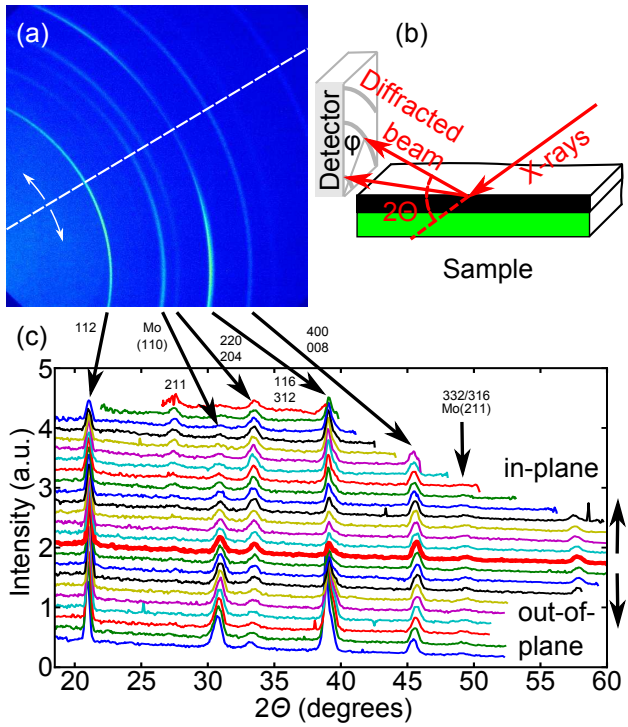}
\caption{\label{meas1} Powder diffraction data. (a) 2D detector image. (b) Powder diffraction pattern integrated along the circular lanes in the 2D image with an integration step of 3$^{\circ}$. The in- and out-of-plane patterns are separated by a thick line at 59$^{\circ}$.}
\end{figure}

An example of a 2D powder diffraction pattern measured on an unetched CIGS-film is shown in Fig.~\ref{meas1}~(a). Measurements geometry is shown in Fig.~\ref{meas1}~(b). The angle $2\theta$ at which the Debye-Scherrer rings occur is defined by the beam energy and a lattice constant of crystallites directed in a specific direction. The intensity change along the circular patterns of the 2D data is associated with the statistical distribution of crystallite orientations and defines a preferred orientation of the crystallites (texture) withing the measured region on the sample~\cite{Wenk2004}. To analyze the distribution the intensity integration is divided into two orientations: the in- and out-of-plane, separated at  $\phi$=59$^{\circ}$, as indicated by the dashed line in Fig.~\ref{meas1}~(a).

In Fig.~\ref{meas1}~(c) a waterfall plot of the data is shown. Every curve is an integration over 3$^{\circ}$ of the angle $\phi$ along the circular patterns. The curves are vertically offset for clarity. The peaks are identified and related to the corresponding peaks in Fig.~\ref{meas1}~(a), as indicated by arrows. This compact data presentation reveals details of the peak evolution depending on the diffraction angle $\phi$. From the peak evolution, we can conclude that the two CIGS peaks with Miller indices (211) and (220)/(204) and the Mo (110) peak have opposite preferred angular orientation dependencies. The bold line separates the in- and out-of-plane crystalline orientations. The ratio of the two integrated intensities is proportional to the ratio of crystallites with different orientation.

Previous measurements of the preferred orientation of CIGS crystallites using the Lotgering factor~\cite{Schlenker2005} showed that the (112) surface orientation is energetically preferred and depends on the Mo substrate. In our experiments, however,  CIGS (112) orientation is under an angle to the sample surface, with out-of-plane orientation predominating. The preferred orientation of Mo (110) plane points out of the sample plane. The preferred orientation of CIGS (116)/(312) plane is slightly out-of-plane, whereas the (211) and (220)/(204) planes are preferably oriented in-plane of the sample. It is important to point out that the preferred orientation of the (211) and (220)/(204) planes are sample dependent.

\begin{figure}
\includegraphics{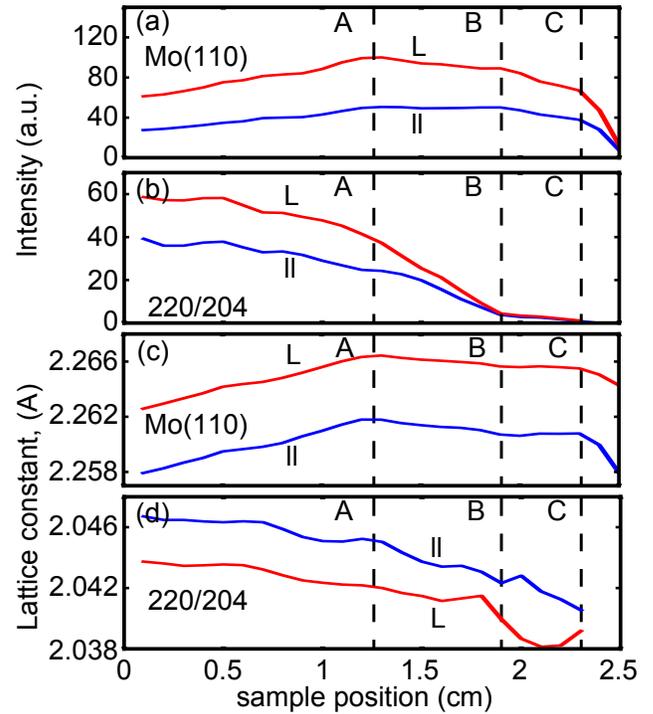}
\caption{\label{meas2} Profiles of X-ray diffraction peak intensities ((a) and (b)) and lattice constants ((c) and (d)) for Mo (110) and CIGS (220)/(204) planes. The data is shown for in-plane (II) and out-of-plane (L) preferred crystalline orientations.}
\end{figure}

2D diffraction images are collected along the etched profile starting from a non-etched region until the layers are etched away. The integrated data for two crystallite orientations are fitted with the Voigt profile. The Mo (110) diffraction peak intensity over the sample is shown in Fig.~\ref{meas2}~(a). As the Mo back contact is covered by the CIGS absorber layer, absorption of X-rays in CIGS decreases the Mo peak intensity. The peak intensity in the experiment increases as the CIGS layer is thinned by etching until position~A (marked by a dashed line in Fig.~\ref{meas2}~(a)) at 1.25~cm on the x coordinate. After this position, the Mo (110) peak intensity decreases, indicating etch pit formation in CIGS layer and removal of the Mo layer. A stronger decrease in the Mo (110) peak intensity is observed after position B at 1.9~cm on the x coordinate. This is associated with the first of the two crystallite layers in Mo (see Figs.~\ref{sample}~(a) and (b)) being removed under the etch pits in CIGS layer. Further Mo (110) peak intensity decrease can be seen after position C at 2.4~cm on the x coordinate and is consistent with removal of the first Mo crystallite layer and direct etching of the second layer. The Mo layer has a ratio of 2 between the intensities of out-of-plane marked by L and in-plane marked by II diffractions for the (110) crystalline orientation.

In Fig.~\ref{meas2}~(b), the CIGS (220)/(204) peak intensity is shown versus the position on the sample. As it is expected, an initial decrease in the peak intensity is caused by thinning of the etched CIGS layer. The etching speed is changed after etch pit formation in the CIGS layer at the position~A. A strong decrease in the CIGS etching speed is observed at position~B and may be caused by the etching reaction that is influenced by an increase in the Mo etching rate, as it has already been explained above. The CIGS (220)/(204) peak intensity is zero after the position~C, indicating that the layer is removed. The ratio between the intensities of out-of-plane and in-plane CIGS (220)/(204) crystalline orientation is 1.6.

In Fig.~\ref{meas2}~(c) lattice constant profiles for the in- and out-of-plane Mo (110) plane are shown. There is a nearly constant offset between the in- and the out-of-plane lattice constant profiles, indicating that this strain anisotropy originates from the substrate. The lattice constants increase with the etching of the CIGS layer, indicating that the Mo lattice is compressed by the strain of CIGS. After position~A, when the top crystallite layer of Mo is etched trough the etch pits in the CIGS layer, the lattice constant decreases to the value defined by the substrate. After position~B, when the first crystallite layer of Mo does no longer play any relevant role, the lattice constant is dominated by the substrate. After position~C, when the bottom crystallite layer of Mo is etched, the Mo (110) lattice constant rapidly relaxes. Lattice relaxation of the layer after position~C is dominated by the in-plane component.

Figure~\ref{meas2}~(d) presents the profiles of CIGS (220)/(204) lattice constants for the in- and out-of-plane crystallite orientations. The lattice is somewhat relaxed in the in-plane direction. It can be seen that the lattice constants decrease over all regions of the etched sample. From the continuous lattice constants decrease, it can be concluded that in the configuration normally used in the thin-film solar cells, namely with a CIGS absorber layer thickness of a few $\mu m$, the absorber layer is not completely relaxed and is influenced by the strain of the Mo substrate. This finding is important for optimization of the solar cell fabrication processes, such as the absorber growth technique.

A confocal scanning microscope was used to perform photoluminescence (PL) spectroscopy measurements. A green (532~nm) laser was applied for PL excitation and an InGaAs detector array near-infrared spectrometer served for the detection of the PL signal.

\begin{figure}
\includegraphics{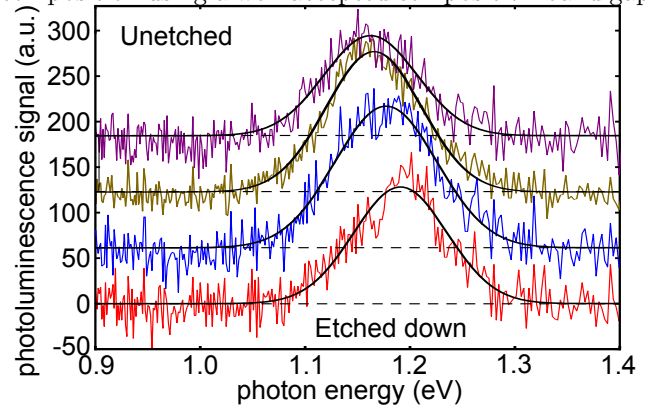}
\caption{\label{meas3} Photoluminescence measurements at different depths in the absorber.}
\end{figure}

In Fig.~\ref{meas3} we present data from spectral PL measurements performed at different depths in the CIGS absorber layer. The curves are offset for clarity. By following the PL peak positions, a band gap profile in the CIGS layer can be estimated. From the total change in the band gap, we calculate an expected change in the composition using a well-accepted composition-band gap dependence~\cite{Dimmler1987} and a bowing factor b=0.12. From the change in the composition, we calculate the change in the lattice constants a0 and c0~\cite{Han2005}. We find that the calculated variation of the CIGS (220)/(204) lattice constants in the absorber layer, caused by the band gap grading, is nearly twice smaller than the variation of the measured lattice constant. From this we conclude that the band gap grading influences the strain distribution within the absorber layer and should be taken into account during growth optimization.

Detailed analysis of strain profiles in the depth of the CIGS absorber is presented in this manuscript. We correlate CIGS (220)/(204) and Mo (110) lattice constant profiles measured over a gradient-etched sample and reveal the influence of the strain induced by the Mo substrate. From the correlation of space-resolved PL and strain profiles, we estimate the influence of the band gap grading. Additionally, we analyze the preferred crystallite orientation in the layers and stress anisotropies.

We gratefully acknowledge financial support from the Concept for the Future in the Excellence Initiative at KIT.

\end{document}